\documentclass[%
 reprint,
superscriptaddress,
 amsmath,amssymb,
 aps,
prb,
floatfix,
]{revtex4-1}

\usepackage{color}
\usepackage{graphicx}% Include figure files
\usepackage{dcolumn}% Align table columns on decimal point
\usepackage{bm}% bold math
\newcommand{\bamnov}{Ba$_{3}$(Mn$_{1-x}$V$_{x}$)$_{2}$O$_{8}$}
\newcommand{\bamnovspace}{Ba$_{3}$(Mn$_{1-x}$V$_{x}$)$_{2}$O$_{8}$ }

\newcommand{\comment}[1]{}

\begin{document}

%\preprint{}

\title{Spin Freezing in The Frustrated Disordered Quantum Magnet \bamnov}% Force line breaks with \\

\author{A. T. Hristov}
\email{hristov@stanford.edu}
\affiliation{
Geballe  Laboratory  for  Advanced  Materials, Stanford  University,  Stanford,  California  94305,  USA
}
\affiliation{
Department  of  Physics, Stanford  University,  Stanford,  California  94305,  USA
}

\author{M. C. Shapiro}
\author{Harlyn J. Silverstein}
\affiliation{
Geballe  Laboratory  for  Advanced  Materials, Stanford  University,  Stanford,  California  94305,  USA
}
\affiliation{
Department  of  Applied Physics, Stanford  University,  Stanford,  California  94305,  USA
}

\author{Minseong Lee}
\affiliation{
Department of Physics, Ulsan National Institute of Science and Technology (UNIST), Ulsan 689-798, Korea
}
\affiliation{%
National High Magnetic Field Laboratory, Florida State University, Tallahassee, Florida 32310, USA
}%
\author{Eun Sang Choi}
\author{L. Rodenbach}
\author{Ju-Hyun Park}
\affiliation{%
National High Magnetic Field Laboratory, Florida State University, Tallahassee, Florida 32310, USA
}%
\author{T. J. S. Munsie}
\affiliation{%
Department of Physics and Astronomy, McMaster University, Hamilton, ON, L8S 4M1, Canada
}%
\author{G. M. Luke}
\affiliation{%
Department of Physics and Astronomy, McMaster University, Hamilton, ON, L8S 4M1, Canada
}
\affiliation{%
Canadian Institute for Advanced Research, Toronto, Ontario M5G 1Z7, Canada
}
\author{L. Civale}
\affiliation{%
Condensed Matter and Magnet Science,
Los Alamos National Laboratory,
Los Alamos, NM 87545 USA
}

\author{I. R. Fisher}
\affiliation{
Geballe  Laboratory  for  Advanced  Materials, Stanford  University,  Stanford,  California  94305,  USA
}
\affiliation{
Department  of  Applied Physics, Stanford  University,  Stanford,  California  94305,  USA
}

\date{\today}% It is always \today, today,

\begin{abstract}
Ba$_3$Mn$_2$O$_8$ is a geometrically frustrated spin dimer compound. 
We investigate the effect of site disorder on the zero field phase diagram of this material by considering the solid solution \bamnov, where nonmagnetic V$^{5+}$ ions partially substitute magnetic Mn$^{5+}$ ions. 
This substitution results in unpaired $S=1$ moments for half-substituted dimers, which are ungapped and therefore susceptible to types of magnetic order not present in the parent compound. 
AC susceptibility measurements of compositions between $x=0.046$ and $x=0.84$ show a sharp frequency- and composition-dependent kink at temperatures below 210mK, suggesting that unpaired spins form a spin glass. 
The case for a glassy state is made clearer by the absence
of any sharp features in the specific heat. 
However, \bamnovspace is not a paradigmatic spin glass. 
Whereas both the freezing temperature and the Weiss temperature (determined from susceptibility above 1K) vary strongly as a function of composition, the heat capacity per unpaired spin is found to be insensitive (above the glass transition) to the density of unpaired spins for the broad regime $0.18\leq x \leq 0.84$. 
This surprising result is consistent with a scenario in which nearest-neighbor unpaired spins form local, possibly fluctuating, spin-singlets prior to the eventual spin freezing. 
The spin glass state is only found for temperatures below the energy scale of single-ion anisotropy, suggestive this plays a significant role in determining the eventual ground state. 
Possible ground states in the ``dilute'' limit ($x < 0.04$ and $x > 0.9$) are also discussed. 
\end{abstract}

\maketitle
%\tableofcontents

\section{\label{sec:level1}Introduction}

Disordered materials realize states of matter that do not exist in the clean limit. 
A particularly rich arena in which to observe and investigate such novel states is provided by quantum spin dimer systems, which are materials characterized by pairs (dimers) of spins with strong antiferromagnetic coupling and weaker inter-dimer interactions.\cite{Zapf_Bose_2014}
At low temperatures, the application of a magnetic field can tune the energies of bosonic excitations, called triplons, in these materials and allow for their condensation into a canted-antiferromagnetic state. \cite{Giamarchi_Bose_2008}
As an example of the effects of disorder in such systems, random bond disorder, which can be achieved by disordering interdimer superexchange interactions via chemical substitution on nonmagnetic sites, localizes triplons into a Bose-glass phase.\cite{ZheludevReview}
In the spin-dimer compounds 
Tl${}_{1-x}$K${}_{x}$CuCl${}_{3}$ 
\cite{TlkCuCl_Shindo2004,TlKCuCl_Suzuki_2009,TlKCuCl_Suzuki_2010,TlKCuCl_Yamada_2011} 
and 
$\mathrm{IPA}\mathrm{\text{-}}\mathrm{Cu}({\mathrm{Cl}}_{x}{\mathrm{Br}}_{1-x}{)}_{3}$ \cite{IPA_Saito1,IPA_ADACHI,IPA_PhysRevLett.101.077204,IPA_PhysRevB.79.092401,Hong_IPA} 
nonmagnetic substitution has allowed for investigation of the Bose-glass to Bose-Einstein condensate (BEC) transition as a function of magnetic field. Such an effect has also been deduced for the closely related case of spin 1 ions with single ion anisotropy.\cite{yu2016bose}

Site disorder, achieved by chemical substitution of nonmagnetic ions on the magnetic sites of spin dimer compounds, offers a complementary means for exploring the nature of the magnetic ground state in the presence of disorder. 
Like random bond disorder, site disorder is anticipated to localize triplons into a glassy state for temperatures and fields close to the BEC phase of the pure dimer system. \cite{Roscilde_Quantum_2005}
However, substitution of nonmagnetic impurities also results in ``unpaired'' magnetic moments on half-substituted dimers \cite{Manna_Tuning_2009} and these spins fundamentally change the important questions that can be asked about the material.
Whereas intact dimers exist in a gapped paramagnetic state and require a magnetic field to induce long range magnetic order, the unpaired spins on broken dimers are not gapped and can order even in the absence of a strong field. \cite{ PhysRevB.84.054417}
The only known experimental result is that, in 1D systems, there is a tendency for unpaired spins to order antiferromagnetically, with the order mediated by excitations of the gapped singlet ground state in unperturbed dimers.
This phenomenon, which is an example of order-by-disorder (OBD), has been observed in studies of nonmagnetic substitution in both the spin-Peierls material Cu$_{1-x}$Zn$_x$GeO$_3$\cite{PhysRevLett.74.1450,Masuda_Phase_1998} and the spin-ladder material Sr(Cu$_{1-x}$M$_x$)$_2$O$_3$\cite{Ohsugi_Impurity_1999} (M=Zn, Ni). 
The results from these quasi-1D systems have been extended to the unfrustrated 2D square lattice by numerical work, which also suggests that an OBD state is achieved for dilute substitution of nonmagnetic impurities.\cite{Roscilde_Quantum_2005}
However, the situation is much less clear for systems with geometric frustration. 

\begin{figure}
\includegraphics[width=.45\textwidth]{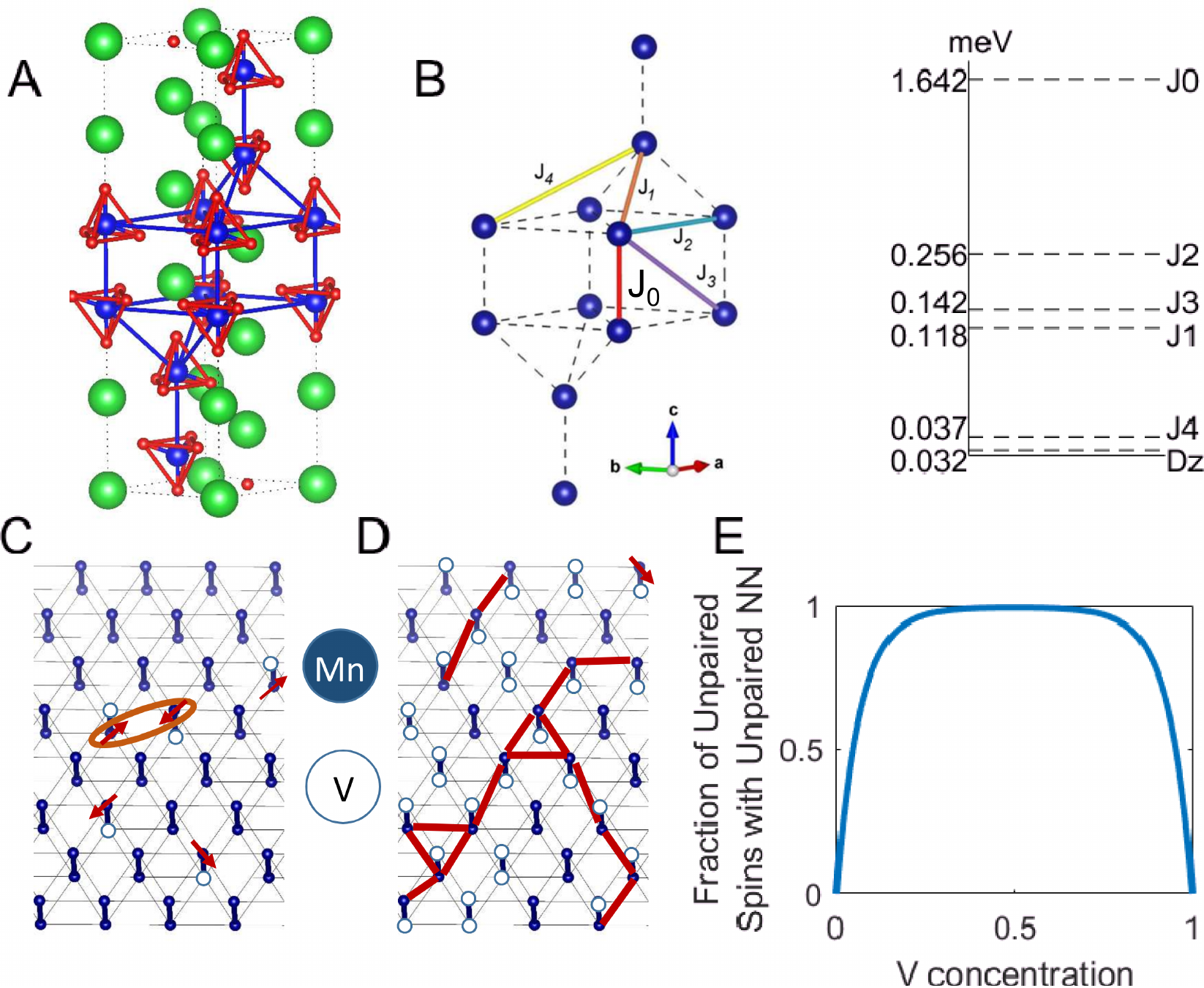}
\caption{
(A) The crystal structure of Ba$_3$Mn$_2$O$_8$ comprises vertical Mn dimers. Each Mn ion (blue) is coordinated by a distorted tetrahedron of oxygen ions (red). Barium ions are shown in green.
(B) Schematic diagram showing the magnetic interactions in Ba$_3$Mn$_2$O$_8$. The diagram shows the same Mn ions as in panel A, but with barium and oxygen ions removed, and solid and dashed lines illustrating the intra-dimer ($J_0$) and interdimer ($J_1$,$J_2$,$J_3$,$J_4$) exchange interactions. 
On the right of the figure, energy values of all exchanges and the uniaxial single ion anisotropy, $D_z$, are shown on a log-scale, revealing the clear separation of energy scales present in this material. 
(C,D) Schematic diagrams of a single bilayer of \bamnov, seen at a small angle from the $c$-axis, revealing the in-plane coordination of vertical dimers. As described in the main text, nonmagnetic V$^{5+}$ ions substitute for magnetic Mn$^{5+}$ ions. In the limit where V substitution is dilute (C), spins from unpaired Mn ions form small clusters (orange) or are isolated, whereas for larger V concentrations (D), networks of unpaired Mn spins (red) grow large and include the majority of unpaired spins.
(E) The fraction of unpaired spins with at least
one unpaired spin in an adjacent unit cell as a function of V concentration $x$. For a wide range of compositions, nearly all unpaired spins have at least one nearest neighbor unpaired spin on an adjacent broken dimer. 
}
\label{fig:crystalStructure}
\end{figure}

In lattices with geometric frustration, all pairwise interactions cannot be satisfied simultaneously. 
In non-dimerized systems, geometric frustration can lead to various types of novel ground states, including spin-liquids\cite{Balents_Nature} and spin-glasses \cite{mydosh}. 
For site-substituted quantum spin dimers on a frustrated lattice, the ground state has not been well established. 
Due to both theoretical challenges (the sign problem, which precludes quantum Monte Carlo studies) and experimental limitations (few candidate materials have been identified), the phase space of possible ground states remains essentially unexplored. 
In this work, we identify one such system (\bamnov) and find, through a series of thermodynamic measurements, that its ground state shows signatures of a spin glass state over a wide range of concentrations ($x$) of the non-magnetic ions. 

Ba$_3$Mn$_2$O$_8$ is an insulator with a large charge gap. 
It crystallizes in the rhombohedral (R$\bar{3}$m) space group and comprises pairs of S = 1 Mn$^{5+}$ ions arranged vertically in triangular planes,\cite{Weller_Ba3Mn2O8_1999,Uchida2002,Tsujii_Specific_2005} which are then stacked with offset to form the rhombohedral crystal structure illustrated in Figure \ref{fig:crystalStructure}.
Distortions of the oxygen tetrahedra around each Mn ion give rise to a uniaxial single ion anisotropy $D_z=-0.032$meV which reduces the energy of the $S_z=\pm 1$ states.\cite{S_Hill_Dz} In isolation, each dimer has a singlet ground state with excited triplet states and a further quintuplet manifold. 
Interdimer exchange interactions, determined from inelastic neutron scattering experiments, are nearly an order of magnitude smaller than the intra-dimer exchanges, leading to a zero-field collective ground state that is a gapped paramagnet with no long range magnetic order.\cite{Stone_Dispersive_2008,PhysRevLett.100.237201} 
Field-induced triplet and quintuplet condensates have been investigated via both thermodynamic and diffraction experiments.\cite{Samulon2008,PhysRevLett.103.047202,Suh_Critical_2009,Samulon_Anisotropic_2010}
Nonmagnetic impurities can be introduced to the dimer lattice by isovalent substitution of nonmagnetic V$^{5+}$ ions for magnetic Mn$^{5+}$, leading to the substitutional alloy series \bamnov. 
Previous work for dilute V substitution ($x < 0.05$) found no indication of the OBD state in the heat capacity down to 50mK and instead suggested that the magnetic ground state in these concentrations was a random singlet phase.\cite{PhysRevB.84.054417}
The present work extends previous studies beyond the dilute limit, revealing clear evidence for an unconventional spin glass in which the removal of entropy in unpaired spins upon cooling is independent of their density, the Weiss temperature extracted from high temperature susceptibility, or even the freezing temperature of the spin glass state into which they eventually enter.

\section{\label{sec:level2}Experimental Methods}
Single crystals of \bamnovspace were grown from a NaOH flux using a phase-pure polycrystalline precursor, as described previously,\cite{Samulon2008} with compositions spanning the entire alloy series (i.e., $0 \leq x \leq 1$).
Compositions were measured for all batches by electron microprobe analysis using Ba$_{3}$Mn$_{2}$O$_{8}$ and Ba$_{3}$V$_{2}$O$_{8}$ standards. 
Samples used in this study were $x$ = 0, 0.02$\pm$0.005, 0.05$\pm$0.02, 0.098$\pm$0.02, 0.18$\pm$0.03, 0.28$\pm$0.02, 0.35$\pm$0.02, 0.45$\pm$0.01, 0.55$\pm$0.03, 0.65$\pm$0.01, 0.84$\pm$0.01, 0.98$\pm$0.006, and 1.0.  The confidence intervals were obtained from measurements of multiple spots on the same crystal. 

Low-field DC magnetic susceptibility measurements were performed with magnetic field along the crystalline $c$-axis in the range between 0.5K and 300K using commercially available systems from Quantum Design. 
Magnetic measurements below 1.8K used the i-Quantum option on instruments at Los Alamos National Laboratory and McMaster University.
For temperatures between 30mK and 700mK, crystals were stacked along their $c$-axis and mounted inside of a purpose-built set of mutual inductance coils, which are sensitive to changes in the real and imaginary components of the AC susceptibility via proportional changes in the mutual inductance. 
The measurements were performed with the entire susceptometer immersed in the mixing chamber of a dilution refrigerator at the National High Magnetic Field Laboratory. 
Due to variations in the volumetric fill fraction of the coils for each composition, quantitative measurements of the susceptibility were only possible for samples where the DC susceptibility has already been measured below the maximum temperature of the dilution refrigerator (approximately 750mK). 
In all cases though, the temperature dependence of the real and imaginary parts of the susceptibility was determined.
Temperatures were extracted from a thermometer mounted near the susceptometer coils, which was previously calibrated for use in this dilution refrigerator. 
Where possible, data were taken while both warming and cooling the sample to account for thermal hysteresis. 
Traces presented derive from measurements where the sample was warming from base temperature, but temperatures extracted from features in the data were obtained from the average of the warming and cooling measurements. 
Changing the sweep rates between 3mK/min and 9.5mK/min yielded no difference in the freezing temperatures extracted by this method.

Heat capacity measurements were performed down to 50mK using relaxation calorimetry in a commercially available Quantum Design Physical Property Measurement System.
Extra care was taken for samples with intermediate concentrations of V due to the large low-temperature heat capacity. 
We found that samples of masses around 0.2mg worked best in this regime on our instrument.
Sample masses were measured using a Mettler Toledo AG245 precision balance, with a typical uncertainty of 0.01mg.

\section{\label{sec:level3}Results}

Measurements performed in this work fall into three categories. 
First, magnetic susceptibility measurements are used to study the structure of quenched randomness in \bamnov.
Results from these first measurements are suggestive that V substituted sites are spatially uncorrelated.
Second, low-temperature susceptibility measurements are used to reveal the formation of a spin-glass magnetic ground state for $0.048\leq x\leq 0.84$. 
Finally, heat capacity measurements are used to show that spin correlations onset at temperatures up to an order of magnitude above the freezing temperature and, for a broad region spanning $0.18 \leq x \leq 0.84$, the entropy per unpaired spin is essentially independent of composition (i.e., apparently does not depend on the number of neighboring spins) even though the eventual glass transition temperature varies with composition.

\subsection{\label{sec:level3a} Concentration of unpaired spins and dimers}

Magnetic states are sensitive not only to the presence of nonmagnetic impurities but also to their spatial distribution. 
In the absence of structural information for Mn or V clustering, we start with a model in which V ions are independently distributed. 
In this model, the probability that any Mn site has been substituted is always $x$ (the stoichiometric concentration of V in the chemical formula \bamnov); from this, it follows that the probability of an intact dimer is $(1-x)^2$ and the probability of only a single spin on a dimer site is $2x(1-x)$. 

The magnetic susceptibility is a powerful tool for investigating the distribution of V impurities because the temperature dependence of the magnetic susceptibility is drastically different between intact dimers and unpaired spins. 
In particular, below the energy scale of the intra-dimer exchange, the susceptibility of an intact dimer decreases with decreasing temperatures, following an exponential behavior at low enough temperatures.
\footnote{
In the case of interacting dimers, which broaden the triplet excitations of an isolated dimer into a band, the spin gap is a not given by $J_0$ but is also a function of the interdimer exchange.
}
 
However, the temperature dependence of unpaired spins continues to rise as temperature is reduced as long as the system does not order or freeze.
Neglecting renormalization effects, a first approximation to the susceptibility can therefore be provided by a linear sum of terms describing the susceptibility of intact dimers ($\chi_d$) and a term describing the Curie-Weiss behavior of unpaired spins\cite{Uchida2002,Samulon2008}:
\begin{eqnarray}
\chi & = & (1-x)^{2} \frac{\chi_{d}}{1+\lambda\chi_{d}} + 2x(1-x)\frac{C_0}{T-\theta} +\chi_{0}, \label{eq:magsuscep1}\\ 
\chi_d & = & \frac{2 N_A \beta g^2 \mu^{2}_{B}(1+5e^{-2\beta J_{0}})}{3+e^{\beta J_{0}}+5e^{-2\beta J_{0}}} \label{eq:magsuscep2}.
\end{eqnarray}
The only free parameters in \eqref{eq:magsuscep1} are the temperature independent background $\chi_0$, the Weiss temperature $\theta$, and a mean-field correction $\lambda$ ($ = \langle \frac{1}{2}\sum_{NN} J' \rangle / N_{A} g^2 \mu_{B}^{2}$, with the sum over nearest neighbor exchanges) to account for interdimer exchange; the Curie constant $C_0$ is determined from fundamental constants and measurements of the $g$-factor while $x$ is measured by electron microprobe analysis.  

\begin{figure}
\includegraphics[width=.45\textwidth]{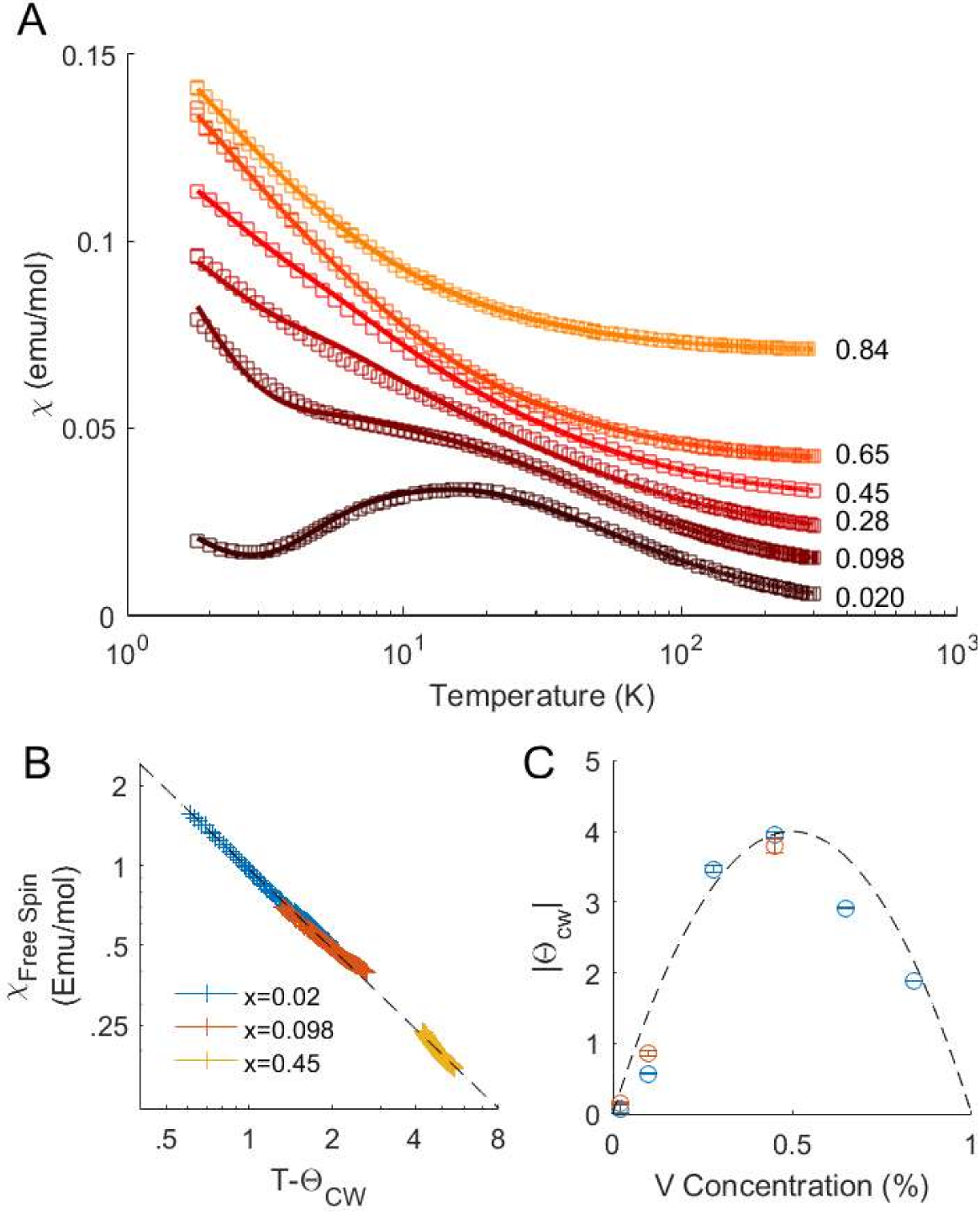}
\caption{(A) The DC magnetic susceptibility of six compositions of \bamnov, offset for clarity. Individual data points correspond to measured values of the susceptibility for each composition; superposed lines derive from model fits to the data (to the form in Eq. \eqref{eq:magsuscep1}), with the Weiss temperature $\theta_{CW}$, temperature independent term $\chi_0$ and the mean-field interaction parameter $\lambda$ as the only free parameters.
Small systematic deviations for intermediate temperatures are ascribed to limitations of the mean-field model used to describe the susceptibility of intact dimers. 
Given the simplicity of the model and the small number of free parameters, the fit is remarkably good, providing strong evidence that the V substituted sites are uncorrelated in the substitutional alloy.
(B) 
DC susceptibilities below 2K for three compositions spanning the region $x<0.5$.  Data are re-scaled by the assumed number of unpaired spins for uncorrelated V substitution and plotted as a function of $T-\theta_{CW}$. The dashed line corresponds to the Curie-Weiss law.
(C) The Weiss temperature of unpaired spins as extracted from measurements for $T<2$K (in red) and $1.8K\leq T\leq 300K$ (in blue). The dashed line is proportional to the assumed density of unpaired spins, $2x(1-x)$.
\label{fig:chiDC}}
\end{figure}

The results of this analysis are shown in Figure \ref{fig:chiDC}(A) for six representative compositions spanning the phase diagram for temperatures down to 1.8K. 
The variations in the Weiss temperature and $x$ are sufficient to explain the temperature dependence in all measured susceptibilities due to insensitivity to the specific value of $\lambda$ (fits obtained by this procedure result in uncertainty in $\lambda$ of $\approx 50\%$). 
We interpret this remarkably good fit to the experimental data as evidence that substitution of V on one site in a dimer does not impact the conditional probability of substitution on the other site of the same dimer (which was an initial assumption used to estimate the number of intact dimers and unpaired spins).
Since nearest neighbor sites are closest and therefore most strongly affected by strain, other forms of spatial correlation (e.g., next-nearest neighbor) are presumably negligible as well.

Due to the exponentially small contribution to the susceptibility from intact dimers at very low temperatures ($T \ll J_0$), the susceptibility in the regime below 2K is well approximated by only the contribution from unpaired spins, which (neglecting any renormalization effects) are assumed to follow a Curie-Weiss form.
However, the concentration of unpaired spins varies with chemical composition, so to compare data for different concentrations it is neccessary to scale measurements by the number of unpaired spins.
Figure \ref{fig:chiDC}(B) shows the magnetic susceptibilities per unpaired spin for three compositions ($x$=0.02, 0.098 and 0.45) at temperatures in the range 0.47K to 2K. 
The data have been fit to a Curie-Weiss law and offset by the fitted Weiss temperature to highlight the universal scaling. 
Weiss temperatures extracted from both sets of measurements (i.e., for $T > 1.8$K in He-4, and $0.47$K $< T <$ $2$K in pumped He-3) are in broad agreement and generally follow the concentration of unpaired spins $2x(1-x)$.
Despite measuring well below the exchange energies between unpaired spins ($J_1$, $J_2$, and $J_3$), no overwhelming deviations from the Curie-Weiss law are observed, even for concentrations where a random distribution of V impurities would lead to most unpaired spins having at least one nearest neighbor exchange to another unpaired spin. 
Indeed, the absence of any signatures of magnetic order necessitated further measurement at lower temperatures to elucidate the magnetic ground state.

\subsection{\label{sec:level3b} Evidence for a Spin Glass State}

\begin{figure*}[t]
\includegraphics[width=0.85\textwidth]{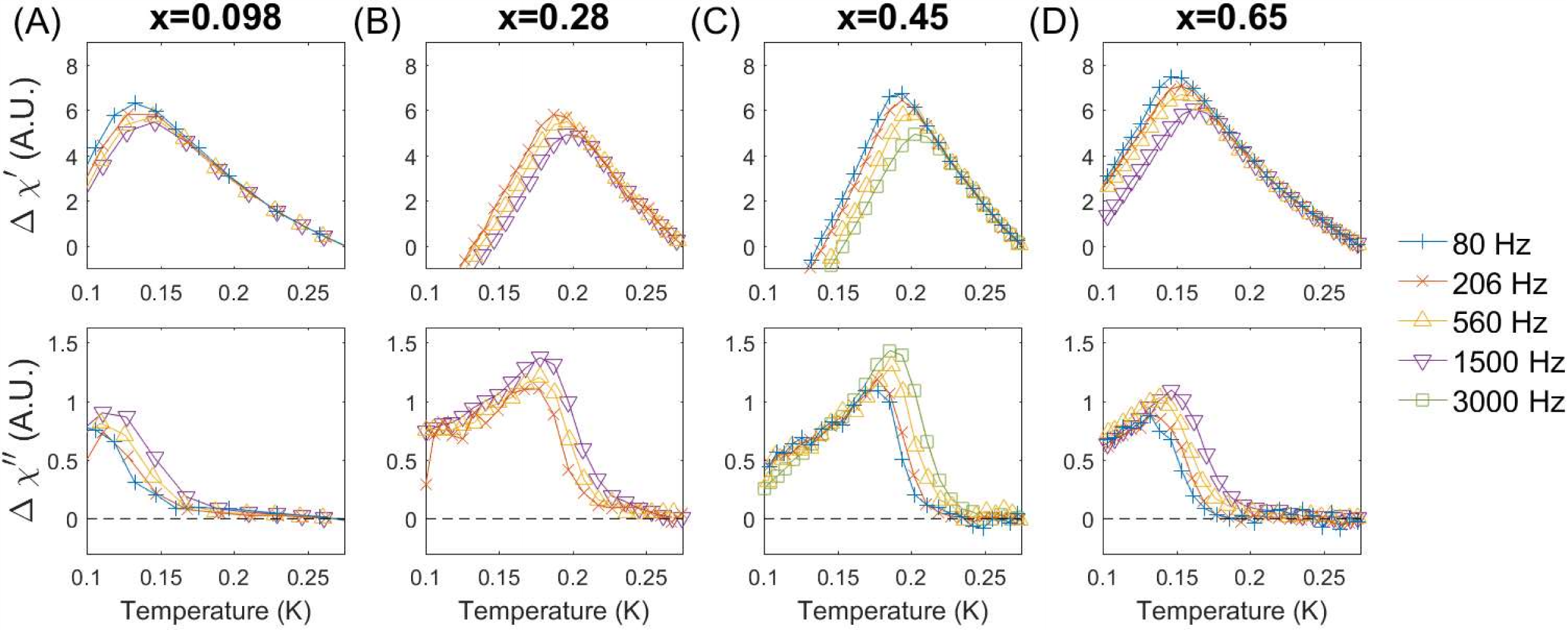}
\caption{Frequency dependence of the real $\Delta \chi'$ (top panels) and imaginary $\Delta \chi''$ (bottom panels) components of the AC magnetic susceptibility of \bamnovspace: $x$ = (a) 0.098, (b) 0.28, (c) 0.45, and (d) 0.65, respectively, as measured relative to the values at $T=275$mK. The frequency dependence of the maximum of $\Delta \chi'$ and the sudden onset of a dissipative imaginary component are both characteristic signatures of spin freezing at the glass transition. Differences in the magnitude of $\Delta \chi'$ and $\Delta \chi''$ between samples are artifacts due to measuring in different susceptometer coils.
\label{fig:freezing}}
\end{figure*}

Representative low temperature AC susceptibility measurements are shown in Figure \ref{fig:freezing} for four compositions in the range $0.098 \leq x \leq 0.66$. 
All of the measurements reveal a sharp kink in the real part of the susceptibility and a jump in the imaginary part of the susceptibility, both of which move to progressively higher temperature as frequency is increased, characteristic of freezing into a spin glass state
\footnote{
This pattern cannot be accounted for by self-heating, which has been ruled out by variation in the excitation current and which would move the peak in the opposite direction.
}
The spin glass freezing temperature $T_f$, as determined from maxima in $\Delta \chi'$ and the non-zero onset of $\Delta \chi''$,  varies with composition and obtains a maximum value in the zero frequency limit of 190$\pm$10mK for compositions close to $x=0.40$.
Additional compositions $x=0.046$ and $x=0.84$ also show this sharp kink, but the frequency dependence has not been investigated.
Self-heating precludes measurements over a wider range of frequency necessary to fit to an Arrhenius law.

\subsection{\label{sec:level3c} Removal of magnetic entropy}
\begin{figure}
\includegraphics[width=.5\textwidth]{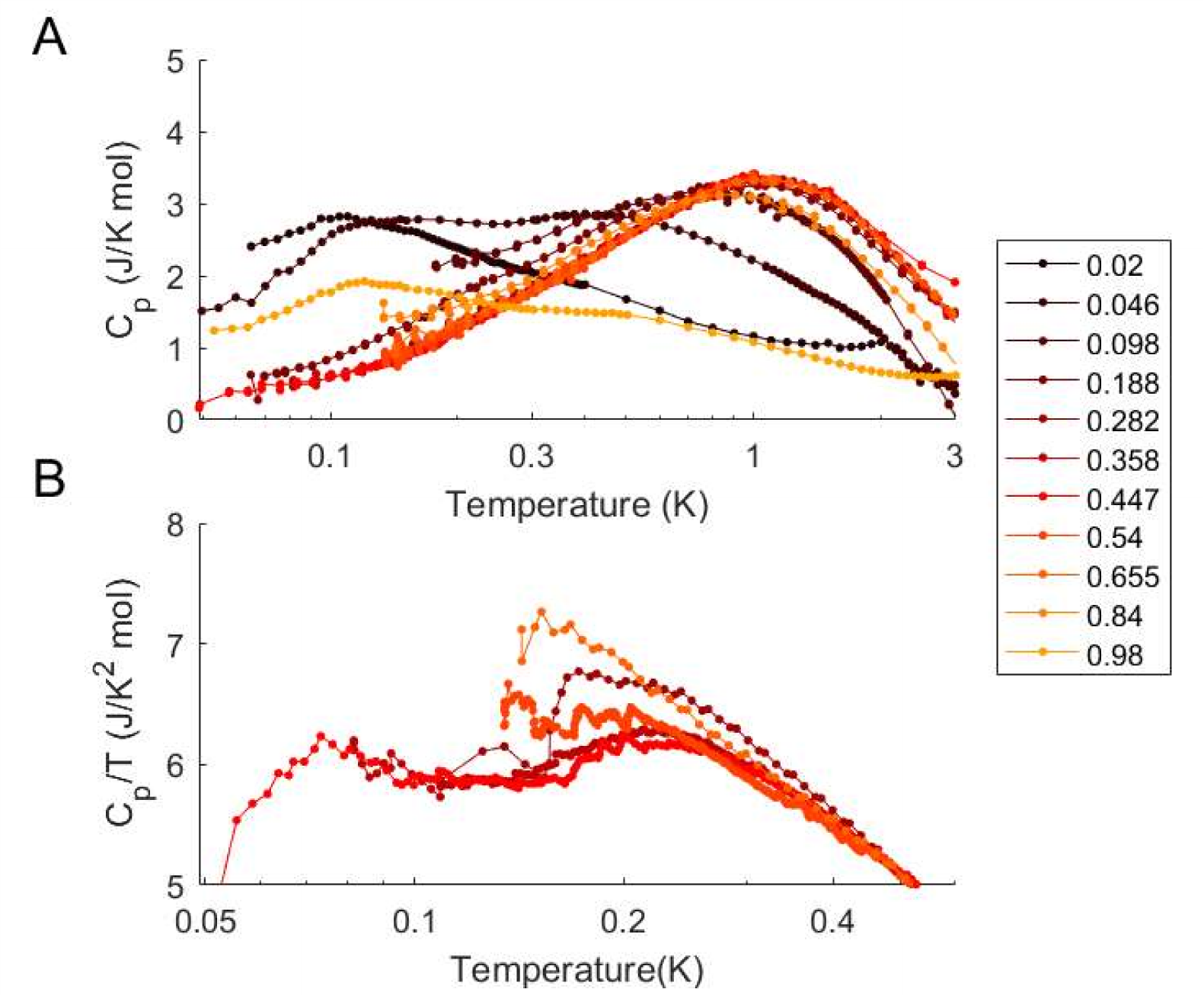}
\caption{(A) Heat capacity per mol of unpaired spins in \bamnovspace, after subtracting the contribution from intact dimers. (B) Heat capacity divided by temperature for compositions $0.28\leq x \leq 0.66$. For $T>300$mK, there is a striking collapse of the data to a universal $T$ dependence.  Small deviations are only visible for temperatures below the energy scale of nearest-neighbor exchanges for unpaired spins.}
\label{fig:heatCapacity}
\end{figure}

Heat capacity measurements were performed to study the removal of entropy per mol of unpaired spins, and the results are shown in Figure \ref{fig:heatCapacity}. 
Having established that V substitution is well described by an independent random distribution across sites, the specific heat of unpaired spins can be isolated from other contributions according to the relation:
\begin{equation}
C_{p}^{\text{unpaired}} = \frac{1}{2x(1-x)}\left(C_{p} - C_p^{\text{phonons}} - (1-x)^2 C_p^{\text{dimer}} \right),
\end{equation}
where $C_p$ is the measured specific heat of a single composition, and the contributions from phonons and dimers are approximated from the Ba$_3$V$_2$O$_8$ and Ba$_3$Mn$_2$O$_8$ standards.
\footnote{
The specific heat of pure Ba$_3$Mn$_2$O$_8$ consists of contributions from phonons, intact triplons, and a small concentration of magnetic impurities, whereas the specific heat of Ba$_3$V$_2$O$_8$ lacks spin dimers. 
For both materials, the heat capacity of magnetic impurities is neglected, as previous measurements have shown its contribution minimal at these temperatures.
Rhetorically, one can suppose the phonon spectra are identical in order to make the approximation $C_p^{dimers}\approx C_p^{\text{Ba}_3\text{Mn}_2\text{O}_8}-C_p^{\text{Ba}_3\text{V}_2\text{O}_8}$ and then to treat $C_p^{phonons}\approx C_p^{\text{Ba}_3\text{V}_2\text{O}_8}$.
After algebraic manipulation, this means the specific of unpaired spins is extracted by taking the measured specific heat less the quantity 
$(1-x)^2C_p^{\text{Ba}_3\text{Mn}_2\text{O}_8}+(2x-x^2)C_p^{\text{Ba}_3\text{V}_2\text{O}_8}$
}
It is known that the dispersion of triplet excitations varies slightly with disorder,\cite{stone_persistence_2011} but since a precise functional form is not known, subtraction based on a pure Mn standard provides a best approximation for the contribution of intact dimers. 
In order to minimize systematic uncertainties arising from such an effect, data were collected and analyzed in the regime where intact dimers are not the dominant part of the heat capacity (below 3K).

Our measurements reveal a bifurcation as a function of composition partitioned approximately where the concentration of unpaired spins is 0.2 (that is, the V concentration is either less than 10\% or greater than 90\%). 
In the region where unpaired spins are more dilute, there is a systematic evolution of the heat capacity with doping and progressively more heat capacity at higher temperatures, indicative of the increasing probability of finding unpaired spins with nearest neighbor exchange interactions as $x$ is increased (Figure \ref{fig:heatCapacity}(A)). 
The more visually stunning aspect of Figure \ref{fig:heatCapacity}(A), however, is the nearly ubiquitous collapse of heat capacity data onto a single universal curve for all compositions in the range $0.188\leq x \leq 0.655$. 
Indeed, Figure \ref{fig:heatCapacity}(B) shows that differences in specific heat between compositions are only evident below 300mK, well below the scale of nearest neighbor interdimer exchanges ($J_1$, $J_2$, and $J_3$).
This pattern is strongly suggestive that the thermodynamic properties are insensitive to the average number of nearest neighbors for each unpaired spin, even though the spin-glass freezing temperature $T_f$ is found to vary with composition.
This remarkable result is in contrast to the behavior of canonical spin glasses, for which entropy is removed at progressively higher temperatures for higher concentrations of magnetic ions.\cite{Wenger_Calorimetric_1976,Martin_Specific_1980}

\begin{figure}
\includegraphics[width=.45\textwidth]{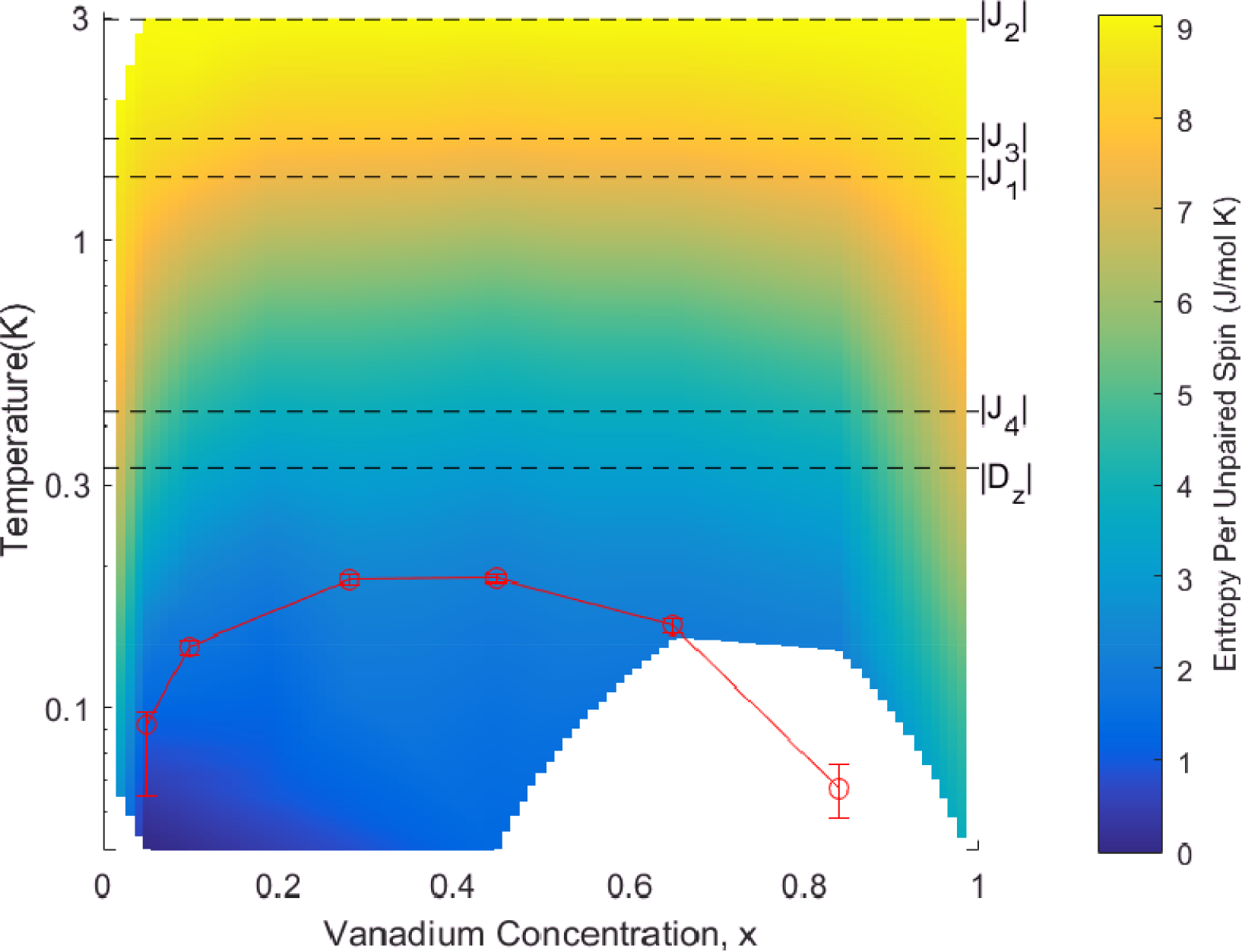}
\caption{
Integrated entropy change per mol of unpaired spins for \bamnovspace as a function of composition, relative to the high temperature value of $R\log(3) = 9.31$J/(mol K) expected from free $S=1$ spins.
Data are shown as a color scale, with a splined interpolation between adjacent compositions. Spin glass transition temperatures are plotted in red.}
\label{fig:Entropy}
\end{figure}

The above result can also be readily appreciated from a direct comparison of the integrated entropy change over temperature as a function of composition.
Entropy is determined from the integral of $C_p/T$, and hence comparison of entropy between samples requires comparing the integrals over the same temperature region (i.e., comparing either $\int_0^T (C_p/T') dT'$ or $\int_T^\infty (C_p/T') dT'$).
Since the heat capacity is appreciable down to the lowest temperatures measured for samples with dilute concentrations of unpaired spins, the low temperature limit is not accessible and thus cannot be used for comparison of the integrated entropy. 
Therefore, entropies are computed relative to the fixed value of $R\log(3)$ expected at high temperature for $S=1$. 
The result, shown in Figure \ref{fig:Entropy} along with the measured glass transition temperatures, shows that upon cooling, approximately 2/3 of the available entropy is already removed before the onset of spin freezing. 
More importantly, a striking uniformity of entropy removal per unpaired spin over the regime $0.188\leq x \leq 0.655$ is evident. For comparison, the glass transition is also plotted in Figure \ref{fig:Entropy}, revealing a clear composition dependence even in the regime where the magnetic entropy per unpaired spin shows minimal variation with $x$.

\section{\label{sec:level4}Discussion}
The existence of a spin-glass ground state somewhere in the phase diagram of \bamnovspace is not alone surprising, as the system exhibits a number of the hallmarks of glassy magnets, including magnetic anisotropy and a combination of ferromagnetic and antiferromagnetic exchanges.\cite{mydosh}
However, there are several noteworthy features across the phase diagram that clearly distinguish \bamnovspace from canonical spin-glass systems.

The first difference to typical spin glass systems is that the freezing temperature does not show a monotonic variation with composition. 
Seen from the perspective of the non-magnetic end-member, as the concentration of Mn "impurities" in Ba$_3$V$_2$O$_8$ increases from zero, $T_f$ rises and then goes through a broad maximum, before decreasing for Mn impurity concentrations beyond approximately 60\% (i.e. for V concentrations x below approximately 40\%). 
This compositional dependence of the freezing temperature reflects the tendency to dimer formation due to strongly antiferromagnetic intra-dimer exchange (relative to the interdimer exchange interactions) favoring local spin singlet states on intact dimers. 
As the concentration of Mn ions increases beyond 50\%, the number of unpaired Mn spins \emph{decreases}, reducing the spin freezing temperature. 
However, despite the fact that the concentration of unpaired spins on half-occupied dimers is symmetric about $x=0.5$, the freezing transition is not. 
Several contributing factors can drive this asymmetry. 
In particular, exchange interactions are not symmetric about $x=0.5$ due to a combination of lattice strain effects (the lattice parameters of Ba$_3$V$_2$O$8$ are different from those of Ba$_3$Mn$_2$O$_8$) and also because for $x \ll 0.5$ unpaired spins interact via intact dimers, which renormalize the exchange interactions \cite{MaDasguptaHu_PhysRevLett.43.1434, Bhatt_Scaling_1982} relative to the ``empty lattice" for $1-x \ll 0.5$. Secondly, changes in the lattice parameters and local strain effects associated with the substitution of V for Mn can induce changes in the single ion anisotropy (from the distortion of the oxygen tetrahedral coordinating each Mn ion), which might vary non-trivially with composition.\cite{[Previous EPR measurements in samples with dilute Mn doping in Ba$_3$V$_2$O$_8$ ($x>0.9$) found Mn$^{4+}$ impurities not present in our samples and a reduced single ion anisotropy of approximately 0.024meV. See ] Whitmore_Electron_1993}

A more striking effect is the near uniformity of the heat capacity and entropy per unpaired spin for all samples in the regime $0.188\leq x \leq 0.84$. 
For the heat capacity to be insensitive to impurity concentration over such a broad region, related quantities should presumably be insensitive to composition as well. 
Remarkably, no other measured quantity has been shown to have a similar composition dependence. 
For example, in a mean-field picture, the sum of exchange interactions weighted by probability of site occupancy sets the Weiss temperature of the magnetic susceptibility.
As the density of unpaired spins changes, so does the expected number of nearest-neighbor unpaired spins, and with it so does the average per-spin sum of exchange interactions.
This relationship is reflected in Figure \ref{fig:chiDC}(D), in which the measured Weiss temperature is shown to track the assumed density of unpaired spins, $2x(1-x)$, in stark contrast to the compositionally-independent magnetic heat capacity and entropy for the same range of compositions.

However, for the independently distributed V model discussed in \ref{sec:level3a}, it is possible to calculate one significant statistical quantity which displays the correct insensitivity to composition.
Figure \ref{fig:crystalStructure}(E) shows the expected fraction of unpaired spins with another unpaired spin in an adjacent unit cell. 
It is useful to name this quantity the nearest-neighbor-coupled fraction (NNCF) of the unpaired spins.
For the regions where the heat capacity is insensitive to composition, the NNCF is nearly constant and close to 1, which suggests the anomalous heat capacity is concomitant to a lattice in which nearly all spins have at least one nearest neighbor (illustrated schematically in Figure \ref{fig:crystalStructure}(D)).
In the limit of dilute V substitution studied previously,\cite{PhysRevB.84.054417} the heat capacity can be well approximated by a model in which entropy is removed by the formation of spin-singlets between neighboring spins. 
A similar process might go some way to explaining the present phenomena for higher V concentrations.
Figure \ref{fig:crystalStructure}(E) illustrates that, for a wide range of intermediate compositions, virtually all unpaired spins have an unpaired neighbor, so if vanishingly few are left out of singlet formation the entropy per unpaired spin would appear essentially identical.
It is not clear what role factors such as geometric frustration play in causing singlet formation for concentrations in which the unpaired spins form an almost percolating network (Figure \ref{fig:crystalStructure}(D)),
\footnote{
The exact concentration at which percolation occurs for unpaired spins is not known for this dimerized lattice. The complication arises from inter-layer bonds between dimers: only the top site of the bottom dimer layer and the bottom site of the top dimer layer interact. 
If we forbid percolation by such means, we overestimate the percolation threshold by projecting to a 2D triangular lattice which has a percolation limit at 50\% unit cell filling, or $2x(1-x)=0.5$. 
On the other hand, we can underestimate the spin percolation threshold by ignoring the dimer sublayers and projecting to the FCC lattice, which has a percolation threshold of $2x(1-x) \approx 0.2$. 
}
but the empirical observation of an insensitivity to composition of the magnetic entropy per unpaired spin is highly suggestive of such a state or something closely resembling it. 
We emphasize that this behavior is different to cluster formation in typical spin glasses, for which magnetic entropy is removed at progressively higher temperatures as the concentration of magnetic impurities is increased.  

Given that the heat capacity is suggestive that entropy is removed by singlet formation between nearest neighbor spins, it is natural to revisit the observed temperature-dependence of the magnetic susceptibility to find evidence for or against such a scenario. 
The susceptibility for a singlet-forming system with a continuous range of exchange couplings between spins continues to rise as temperature is reduced, but will not generically follow a Curie-Weiss law.\cite{Bhatt_Scaling_1982} 
The susceptibility data shown in Figure \ref{fig:chiDC} were not taken over a wide enough temperature range to distinguish subtle deviations from Curie-Weiss behavior. 
However, given the relatively well-separated energy scales of nearest neighbor exchange interactions in Ba$_3$Mn$_2$O$_8$ (Figure \ref{fig:crystalStructure}(B)),  the apparent absence of any dimer-like maximum in the susceptibility around the equivalent temperature scale is noteworthy. 
Even so, no firm comparison can be made in the absence of calculations that explicitly consider the full range of exchange couplings and employ an appropriate decimation procedure to account for the effect of singlet formation. 
Further measurements to much lower temperatures could empirically address whether  the susceptibility is more accurately expressed by a mean-field antiferromagnetic interaction between spins, or by a decimation procedure which removes degrees of freedom through progressive singlet formation, and hence whether the eventual freezing represents collective degrees of freedom or the spins left unpaired after singlet formation.

The onset of spin glass behavior below the energy scale of the single ion anisotropy $D_z$ is also highly suggestive.
Since $D_z$ serves to lower the energy of the $S_z\pm 1$ states relative to the $S_z=0$ state, cooling across this energy scale represents a crossover to more Ising-like character for the relevant spins. 
It is unclear if the glassiness is solely a consequence of the inherent randomness in the site-substituted lattice, or if it is induced or strongly enhanced by this crossover to Ising behavior. 
Comparison with a spin 1/2 analog, such as Ba$_3$(Cr$_{1-x}$V$_x$)$_2$O$_8$, has the potential to answer this latter question.

For dilute concentrations of unpaired spins ($x < 0.04$ and $x > 0.85$), the probability of unpaired spins having an unpaired nearest neighbor (Figure \ref{fig:crystalStructure}(E)) rapidly drops off on approach to the end members of the alloy series, and magnetic entropy is consequently removed at much lower temperatures (Figure \ref{fig:Entropy}). 
Both the Weiss temperature (Figure \ref{fig:chiDC}(C)) and $T_f$ (Figure \ref{fig:Entropy}) also drop rapidly in this regime, and for sufficiently dilute concentrations we anticipate the glass transition temperature decreases below the base temperature of the present measurement techniques.
Hence, the ground state (or states) in the dilute limit has (have) not been identified in the present work.

For temperatures above the single ion anisotropy energy scale, the cascade of well separated exchange energies makes a random singlet picture a good approximation for the small clusters of coupled spins that apparently form, progressively removing magnetic entropy\cite{PhysRevB.84.054417}. 
However, to rigorously investigate the zero temperature limit, the presence of uniaxial single ion anisotropy cannot be neglected. 
Consequently, this means treating unpaired Mn as Ising spins, for which previous numerical work suggests the distribution of effective exchanges narrows upon cooling, disfavoring the random singlet over other forms of order.\cite{Bhatt_Scaling_1982} 
A Griffiths phase, characterized by rare-yet-large regions of strong magnetic interactions that give rise to non-analytic magnetic behavior, is also possible\cite{Griffiths_Nonanalytic_1969,Bray_Nature_1987} but seems unlikely given that large magnetic regions evident for the concentrated limit ($x\approx0.5$) show no such non-analytic behavior. 
The absence of data to yet lower temperatures means that an order-by-disorder phase\cite{Roscilde_Quantum_2005} cannot be presently ruled out for these compositions.
It is unknown whether either uniaxial anisotropy or geometric frustration might preclude this form of long range order in \bamnov. 
In addition, it is not clear that the character of the magnetic ground state for dilute concentrations should be equivalent for $x \ll 0.5$ (small V concentrations) and $1-x \ll 0.5$ (small Mn concentrations), even though the number of unpaired spins is symmetric about x = 0.5. 
For example, even if an order-by-disorder state is favored as the true ground-state for small V concentrations, the absence of intact dimers in the region $x>0.84$ would make such a state highly unusual for the majority V-substituted samples.
Of course, it is also possible that a spin-glass is formed for all compositions. 
In this case, further comparisons of the temperature scales of spin glass formation and entropy removal could be quite illustrative in answering whether the anomalous region where entropy removal per unpaired spin appears nearly independent of composition is a signature of a more ubiquitous feature of the phase diagram.

\section{\label{sec:level5}Conclusion}

The ground state of a quantum spin dimer magnet is generally not known for systems with geometric frustration and disorder.
Answers to this question are limited both experimentally (due to available materials) and theoretically (due to the frustration-induced sign problem in quantum Monte Carlo simulations).
The present work has investigated this problem in a system, Ba$_3$Mn$_2$O$_8$, with $S=1$ Mn dimers which has been disordered through isovalent substitution of nonmagnetic V for Mn. 
Magnetic susceptibility and heat capacity measurements show that the ground state for much of the alloy series of \bamnovspace is an atypical spin-glass in which entropy removal is largely independent of the spin-glass freezing temperature $T_f$ and onsets at temperatures greater than $T_f$ by more than an order of magnitude.
For the dilute limits where $x<0.04$ or $x>0.90$, the ground state is not presently known, but is likely restricted to further spin-freezing or a highly suppressed order-by-disorder transition. 
These unusual properties merit further investigation into not only the particular material system here discussed, but more generally into the interplay of frustration and disorder in the wider class of quantum magnets.

\section*{acknowledgments}

The authors would like to acknowledge constructive feedback from S.A. Kivelson and R.N. Bhatt and assistance with microprobe analysis from Dale H. Burns and R. E. Jones. 
Work at Stanford was supported by the National Science Foundation, under Grant No. DMR-1205165. 
A.T.H. is supported by a National Science Foundation Graduate Research Fellowship, under Grant No. DGE-114747.
H.J.S. acknowledges support from the Karel Urbanek Fellowship and the National Sciences and Engineering Research Council of Canada.

\bibliography{refs}

\end{document}